\documentclass[twocolumn,showpacs,preprintnumbers,amsmath,amssymb]{revtex4}
\draft
\usepackage{graphicx}
\begin{document}

\title{Spiral-wave Dynamics Depends Sensitively on
Inhomogeneities in Mathematical Models of Ventricular Tissue}
\author{T K Shajahan$^1$}
\author{Sitabhra Sinha$^2$}
\author{Rahul Pandit$^{1,3}$}
\affiliation{$^1$Centre for Condensed Matter Theory,
Department of Physics, Indian Institute of Science,
Bangalore 560012, INDIA\\
$^2$The Institute of Mathematical Sciences, C. I. T. Campus, Taramani, Chennai 600113,  INDIA\\
$^3$also at Jawaharlal Nehru Centre for Advanced Scientific Research,
Bangalore 560064, INDIA}
\begin{abstract}

Every sixth death in industrialized countries occurs because of cardiac
arrhythmias like ventricular tachycardia (VT) and ventricular fibrillation
(VF).  There is growing consensus that VT is associated with an unbroken 
spiral wave of electrical activation on cardiac tissue but VF with broken 
waves, spiral turbulence, spatiotemporal chaos and rapid, irregular activation.
Thus spiral-wave activity in cardiac tissue has been studied extensively.
Nevertheless, many aspects of such spiral dynamics remain elusive because 
of the intrinsically high-dimensional nature of the cardiac-dynamical system.
In particular, the role of tissue heterogeneities in the stability of cardiac
spiral waves is still being investigated. Experiments with conduction inhomogeneities 
in cardiac tissue yield a variety of results: some suggest that conduction inhomogeneities can
eliminate VF partially or completely, leading to VT or quiescence, but
others show that VF is unaffected by obstacles. We propose theoretically that
this variety of results is a natural manifestation of a complex, fractal-like boundary
 that must separate the basins of the attractors associated, respectively, 
with spiral breakup and single spiral wave. We substantiate this with extensive numerical studies of 
Panfilov and Luo-Rudy I models, where we show that the suppression of spiral 
breakup
 depends sensitively on the position, size, and nature of the 
inhomogeneity. 

\end{abstract}
\maketitle
\section{Introduction}
The challenge of understanding the dynamics of spiral waves in excitable 
media is especially important in cardiac tissue where such waves are
implicated in life-threatening arrhythmias such as ventricular tachycardia
(VT) and ventricular fibrillation (VF)\cite{cross,winfree,jalife,gray,witkowsky,garfinkel,gray2}
. {\it Anatomical reentry} because of conduction inhomogeneities in cardiac 
tissue, and {\it functional reentry}\cite{jalife}, which result from wave propagation around transiently inexcitable regions,
are crucial for the initiation of RS (a single rotating spiral wave)
and the initiation and maintenance of ST (spiral turbulence with
broken waves). But the precise ways in which spiral
waves are affected by obstacles in ventricular tissue is 
still not clear\cite{weiss}. Spiral waves form when
waves of excitation circulate around an anatomical obstacle\cite{wiener}. 
However, 
Allesie {\it et al} \cite{allessie} have shown that spiral-wave formation
can also occur with a functionally determined heterogeneity in the tissue.
The interaction of such a wave with an anatomical obstacle can be quite
complex especially in the spatiotemporally chaotic state associated with
spiral turbulence. Indeed, experiments with obstacles in cardiac
tissue have yielded a variety of results. For example, some
experiments\cite{ikeda} report that small obstacles
do not affect spiral waves but, as the size of the obstacle
is increased, such a wave can get pinned to the obstacle.
Various other experiments have discussed the role of an 
anatomical obstacle as an anchoring site for spiral waves,
which can lead to the conversion of ST into RS  \cite{valderrabano,pertsov,kim}.
Davidenko {\it et al} \cite{davidenko} have found that, 
when they induced spiral waves in cardiac
tissue preparations ``... in most episodes, the spiral was anchored to small arteries
or bands of connective tissue, and gave rise to stationary rotations.
In some cases the core drifted away from its site of origin 
and dissipated at the tissue border."   
Other studies have 
shown \cite{valderrabano2,valderrabano3,wu,ohara}
that an obstacle, in the path of a moving spiral wave, can break it and
lead to many competing spiral waves. Recent experiments by Hwang
{\it et al} \cite{hwang} have suggested that multistability of spirals
with different periods in the same cardiac-tissue preparation 
can arise because of the interaction of spiral
tips with small-scale inhomogeneities. 

Conduction inhomogeneities in the ventricle include scar tissues, 
resulting from an infarction, or major blood vessels. Some theoretical
studies of the effects of tissue inhomogeneities have been carried out by
using model equations for cardiac tissue; however, they have not addressed 
the issues we concentrate on. The interaction of an excitation wave with
piecewise linear obstacles has been studied by Starobin {\it et al}
\cite{starobin}  to understand the role of obstacle curvature in the pinning
of such waves.  Xie {\it et al} \cite{xie}  have considered spiral
waves around a circular obstacle and given a plausible connection 
of the ST-RS transition to the size of the 
obstacle. Panfilov {\it et al} \cite{panf-prl,tusscher,tusscher2}  have shown that 
a high concentration of randomly distributed non-excitable cells
can suppress spiral break up. 
Conduction inhomogeneities can also play a very important role in pacing 
termination of cardiac arrhythmias \cite{sinha};
in particular, it is easier to remove a spiral wave once it
is pinned to an obstacle, as described in Refs. \cite{takagi,biktashev}, 
than to control a state with spiral-turbulence.

Here we initiate a study that has been designed specifically to systematize 
the effects of conduction inhomogeneities in mathematical models for 
cardiac arrhythmias. Our work shows clearly that ST can be suppressed or 
not suppressed by obstacles of different sizes depending on
where they are placed. As we argue below, this {\it sensitive dependence} on 
the sizes and positions of obstacles must be a manifestation of 
a complex, fractal-like boundary \cite{lai,sommerer} between the 
domains of attraction of ST and RS. We also show that inhomogeneities
in parameters, which govern ratios of time scales, lead to similar
results.
The models and numerical methods used by us
is described in section II. Section III contains results; and we end with a 
discussion in Section IV.

\section{Models and Numerical Methods}

We use the Panfilov \cite{panf-chaos,panf-pla} and Luo-Rudy I
\cite{luorudy,shajahan} models for cardiac tissue
in our studies; the former is well suited for extensive numerical studies 
because of its relative simplicity; the latter, being realistic, allows 
us to check that the results we obtain are qualitatively correct and not 
artifacts of the Panfilov model.

The Panfilov model 
\cite{panf-chaos,panf-pla} consists of 
two coupled equations, one a partial differential equation (PDE)
and the other an ordinary differential equation (ODE), that specify the 
spatiotemporal evolution of the scaled 
transmembrane potential $V$ 
(denoted by $e$ in Refs. \cite{panf-chaos,panf-pla}) and 
the recovery variable $g$, into which this model lumps all
the effects of the different ion channels: 
\begin{eqnarray}
\label{pan}
\nonumber
{\partial V}/{\partial t} & = & {\nabla}^2
V - f(V) - g; \label{Panfeq}\\ 
{\partial g}/{\partial t} & = & {\epsilon}(V,g) (kV - g).
\end{eqnarray}
The initiation of action potential is encoded in $f(V)$, which
is piecewise linear:
$f(V)= C_1 V$, for $V<e_1$, $f(V) = -C_2 V + a$,
for $e_1 \leq V \leq e_2$, and $f(V) = C_3 (V - 1)$, for $V > e_2$.
The physically appropriate parameters given in
Refs. \cite{panf-chaos,panf-pla} 
are $e_1 = 0.0026$, $e_2 = 0.837$, $C_1 = 20$,
$C_2 = 3$, $C_3 = 15$, $a = 0.06$ and $k = 3$. 
The function $\epsilon (V,g)$ determines the dynamics of the
recovery variable: $\epsilon (V,g) = \epsilon_1$ for
$V < e_2$, $\epsilon (V,g) = \epsilon_2$ for $V > e_2$, and
$\epsilon (V,g) = \epsilon_3$ for $V < e_1$ and $g < g_1$ with
$g_1 = 1.8\/$, $\epsilon_1 = 0.01\/$, $\epsilon_2 = 1.0\/$, and
$\epsilon_3=0.3$.
As in Refs. \cite{panf-chaos,panf-pla}, we define dimensioned time 
$T$ to be 5 ms times dimensionless time and 1 spatial unit to be 1 mm.
The dimensioned value of the conductivity constant $D$ is 2 cm$^2$/s.

In spite of its simplicity, relative to the Luo-Rudy I (LRI) model 
described below, the Panfilov model has been shown to capture several 
essential features of the spatiotemporal evolution of $V$ in cardiac 
tissue \cite{panf-chaos,panf-pla,pandit,shajahan2}. 
As in the LR I model the Panfilov model also contains an absolute and a relative refractory period.
The ways in which spiral patterns appear, propagate and break up,
and the methods by which they can be controlled are very similar in these 
models. 
To make sure that the qualitative features we 
find are not artifacts of the Panfilov model we show explicitly, in 
illustrative cases, that they also occur in the realistic Luo-Rudy I 
model, which is based on the Hodgkin-Huxley formalism and takes into 
account the details of 6 ionic currents (e.g., Na$^+$, K$^+$, and Ca$^{2+}$) 
and 9 gate variables for the voltage-gated ion channels that regulate
the flow of ions across the membrane \cite{luorudy}. 
The concentration difference of the ions, inside and outside
the cell, induces a potential difference
of approximately -84 mV across the cell membrane in 
the quiescent state. Stimuli, which raise the potential across
the cell membrane above -60 mV, change the conductivity 
of the ion channels and yield an action potential
that lasts typically for about 200 ms. Once an action potential is initiated there is
a refractory period during which the same stimulus cannot lead to further 
excitation. Single cells in the Luo-Rudy model are coupled diffusively; thus
one must solve a PDE for the transmembrane 
potential $V$; the time evolution
and $V$ dependence of the currents in this PDE are given by 7 coupled ordinary
differential equations \cite{luorudy,shajahan} which we give in the Appendix.

We integrate the Panfilov model PDEs in $d$ spatial dimensions
by using the forward-Euler method in time $t$, with a
time step $\delta t =0.022$, and a finite-difference method in space, 
with step size  $\delta x =0.5$ and five-point and seven-point 
stencils, respectively, for the Laplacian in $d$=2 and $d$=3. 
Our spatial grids consist of square or simple-cubic lattices with side
$L$ mm, i.e., $( 2 L)^d $ grid points; we have used $L$=200. 
Similarly for the LRI model PDEs we use a forward-Euler method for time
integration, with $\delta t =0.01$ ms, a finite-difference method in space,
with  $\delta x =0.0225$ cm, and a square simulation domain with
$400 \times 400$ grid points, 
i.e., $L$=90 mm. We have checked in representative 
simulations on somewhat smaller domains that a Crank-Nicholson scheme
yields results in agreement with the numerical scheme described above.

\begin{figure*}
\centering
\includegraphics[width=14cm]{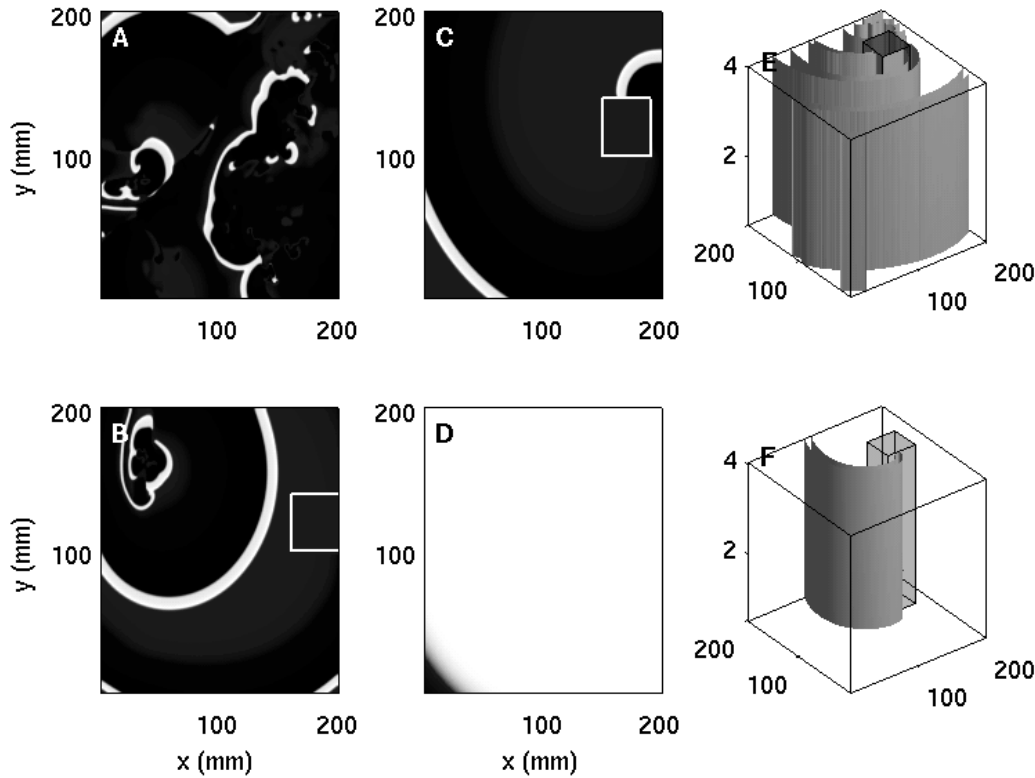}
\caption{
{\bf Panfilov-model spiral turbulence (ST)}: Transmembrane potentials for two dimension 
(pseudo-greyscale plots A-F) and three dimension (isosurface plots G and H).
Two-dimensions:200 mm $\times $200 mm domain and a 40 mm $\times$ 40 mm  
square obstacle with left-bottom corner at ($x, y$).  
({\bf A}) no obstacle -ST;
({\bf B}) ($x=160, y=100$) ST persists;
({\bf c}) ($x=150, y=100$) ST replaced by RS (one rotating anchored spiral);
({\bf D}) ($x=140, y=100$) spiral moves away (medium quiescent).
Three-dimensional analogs of (B) and (C): 
($ 200 \times 200 \times 4$) domain; an obstacle of height 
4 mm and a square base of side 40 mm at ({\bf E}) ($x=140, y=120, z=0$) and
({\bf F}) ($x=140, y=110, z=0$).
}
\end{figure*}

For both models we use no-flux (Neumann) boundary conditions on the edges of
simulation domain and on the boundaries of obstacles. We introduce conduction inhomogeneities in the medium 
by setting the diffusion constant $D$ equal to zero in regions with obstacles;
in all other parts of the simulation domain  $D$ is a nonzero constant. 
The dimensioned value of $D$ is 2 cm$^2$/s for the Panfilov model
and between 0.5 cm$^2$/s and 1 cm$^2$/s for the LRI model; we use 
$D$=0.5 cm$^2$/s in the LRI simulations we report here . In most
of our studies the inhomogeneity is taken to be a square region of
side $l$ , with 10 mm $\leq l \leq 40$ mm;
however, we have also carried out illustrative simulations with
circular or irregularly shaped inhomogeneities. In our three-dimensional
simulations we use an obstacle of height 
4 mm and a square base of side 40 mm, 
i.e., 8 and 80 grid points, respectively (For a detailed understanding 
of the three-dimensional case we must also consider the effects of
rotational anisotropy of muscle fibers in cardiac tissue\cite{fenton}, but
this lies outside the scope of our study.) 
We also study inhomogeneities
in which $\epsilon_1$ in model (1) varies over the simulation domain but $D$ is
constant.

\begin{figure*}
\centering
\includegraphics[width=14cm]{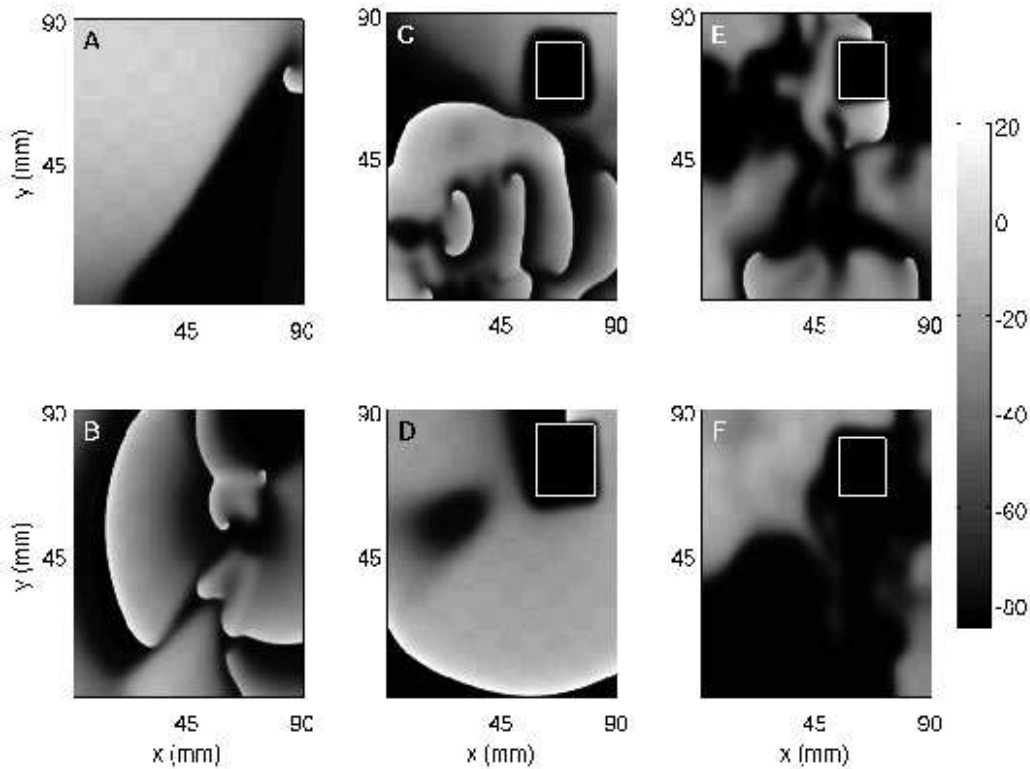}
\caption{
{\bf Luo-Rudy-Model spiral turbulence:} Pseudo-greyscale plots in a
 90 $\times$ 90 mm$^2$ illustrating how the initial condition ({\bf A})
evolves, in the absence of obstacles, to ({\bf B}) via the generation of 
spiral waves and their subsequent breakup. In the presence of a square 
obstacle of side $l$ placed with its bottom-left corner at ($x, y$) 
we obtain the following: ({\bf C})$l$=18 mm and ($x=58.5$ mm, $y=63$ mm) ST persists;
({\bf D})$l$=22.5 mm and ($x=58.5$ mm, $y=63$ mm) RS (one spiral anchored at the obstacle);
for $l$=18 and ($x=54$ mm,$y=63$ mm) spirals disappear leaving the medium quiescent
({\bf E}) at 800 ms and ({\bf F}) at 1000 ms.
}
\end{figure*}
The initial conditions we use are such that, in the absence of inhomogeneities,
they lead to a state that displays spatiotemporal chaos and spiral turbulence.
For the Panfilov model we start with a broken-wavefront initial condition:
For a system of linear size $L$ at time $t$=0 we set $g$=2, for $0 \leq x \leq L$
and $0 \leq y \leq \frac{L}{2}$, and $g = 0$ elsewhere, and $V = 0$ everywhere 
except for $y = \frac{L}{2} + 1$ and $0 \leq x \leq \frac{L}{2}$, where
$V = 0.9$.  From this broken wavefront a spiral wave develops 
with a core in the centre of the simulation
domain and, in the absence of inhomogeneities, evolves to a state with broken
spiral waves and turbulence (Fig. 1A). The spirals continue to break up 
even after 35000 ms for the parameters we use.
For the LRI model we start from the
initial condition shown in Fig. 2A which develops, without an obstacle, into
the spiral-turbulent state shown in Fig. 2B. 

In the presence of an obstacle the spiral turbulence (ST) state of Fig. 1A 
can either remain in the ST state or evolve into a quiescent state (Q)
with no spirals or the RS state with one rotating spiral anchored at 
the obstacle. We explore all these possibilities in the next Section.
Before we do so, we give the criteria we use to decide whether a given 
state, of the system we consider, is of type ST, RS, or Q.
In the Panfilov model, if the spiral wave continue to form and break up even up to 3500 ms, we identify the state as ST (Fig. 1B); if, by contrast, 
a single spiral wave anchors to the obstacle and rotates around it
at least for ten rotations ($\simeq$ 3500 ms for the Panfilov model with
a 40 $\times$ 40 mm$^2$ obstacle) we say that an RS state (Fig. 1C) has 
been achieved (we have seen that, once it anchors, this rotation of 
the spiral wave continues even after 100 rotation periods); lastly, if
the spiral wave moves away from the simulation domain and is absorbed at 
the boundaries within 3500 ms, we conclude that the state is Q (Fig. 1D).

For the LRI model, if the spiral formation and break up continues
upto 2200 ms, we identify the state as ST (Fig. 2C); if the spiral 
wave gets anchored to the obstacle and completes 4 rotation periods
($\simeq$ 2200 ms for the obstacle we use)
we identify the state as RS (Fig. 2D); and we say that the state Q
(Figs. 2E and 2F) is achieved if the spiral wave moves away from the 
simulation domain within 2200 ms.  

\section{Results}
Cardiac tissue can have conduction inhomogeneities at various length scales.
Even minute changes in cell or gap-junctional densities might act as
conduction inhomogeneities\cite{hwang}; these are of the 
order of microns. Scar tissues or blood vessels can 
lead to much bigger obstacles; these are in the mm to
cm range so they can be studied effectively by using the PDEs mentioned 
above. Here we focus on such large obstacles. As in the experiments of Ikeda
{\it et al} \cite{ikeda}, we fix the position of the obstacle and 
study spiral-wave dynamics as a function of the obstacle size.
For this we introduce a square obstacle of side $l$ in the two-dimensional
($d$ = 2) Panfilov model in a square
simulation domain with side $L$=200 mm. We find that, with the bottom-left
corner of the obstacle at the point (50 mm, 100 mm) 
spiral turbulence (ST)
persists if $l \leq (40-\Delta) $ mm, a quiescent state (Q) with no spirals is obtained 
if $l$= 40mm, and a state with a single rotating spiral (RS) anchored at the
obstacle is obtained if $l \geq (40+\Delta) $ mm. To obtain these results
we have varied $l$ from 2 to 80 mm in steps of $\Delta$= 1 mm.
Hence there is a clear transition from 
spiral turbulence to stable spirals, with these two states separated by a
state with no spirals. 
\begin{figure*}
\centering
\includegraphics[width=16cm]{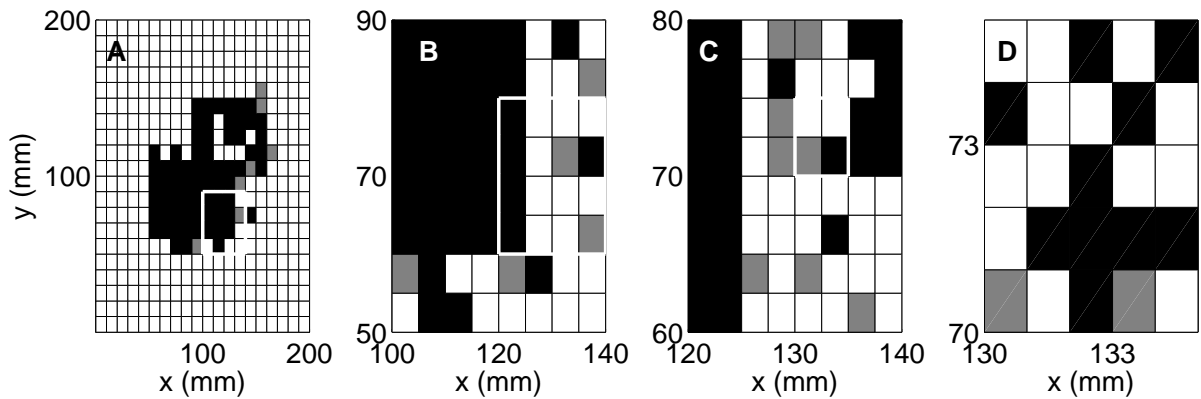}
\caption{
{\bf Panfilov-model stability diagram:} The effect of an 40 $\times$ 40 mm$^2$ 
obstacle in a 200 $\times$ 200 mm$^2$ domain shown by small squares (side $l_p$)
the colors of which indicate the final state of the system when the position
of the bottom-left corner of the obstacle coincides with that of the small
square (white, black, and gray denote ST, RS, and Q, respectively). 
({\bf A}) for $l_p$=10 mm. We get the fractal-like structure of the interfaces
 between ST, RS, and Q by zooming in on small sub domains encompassing 
parts of these interfaces (white boundaries in {\bf A}, {\bf B}, and {\bf C} with
({\bf B}) $l_p$=5 mm, ({\bf C}) $l_p$=2.5 mm, and ({\bf D}) $l_p$= 1 mm).
}
\end{figure*}

The final state of the system depends not just on the size of the obstacle
but also on how it is placed with respect to the tip of the initial wavefront. In our
simulations we find, e.g., that even a small obstacle, placed close to the
tip [$l$=10 mm obstacle placed at (100 mm, 100 mm)], can prevent the 
spiral from breaking up, whereas a bigger obstacle, placed far away from the
tip [$l$= 75 mm, placed at (125 mm, 50 mm)], does not affect the spiral.

\begin{figure*}
\centering
\includegraphics[width=14cm]{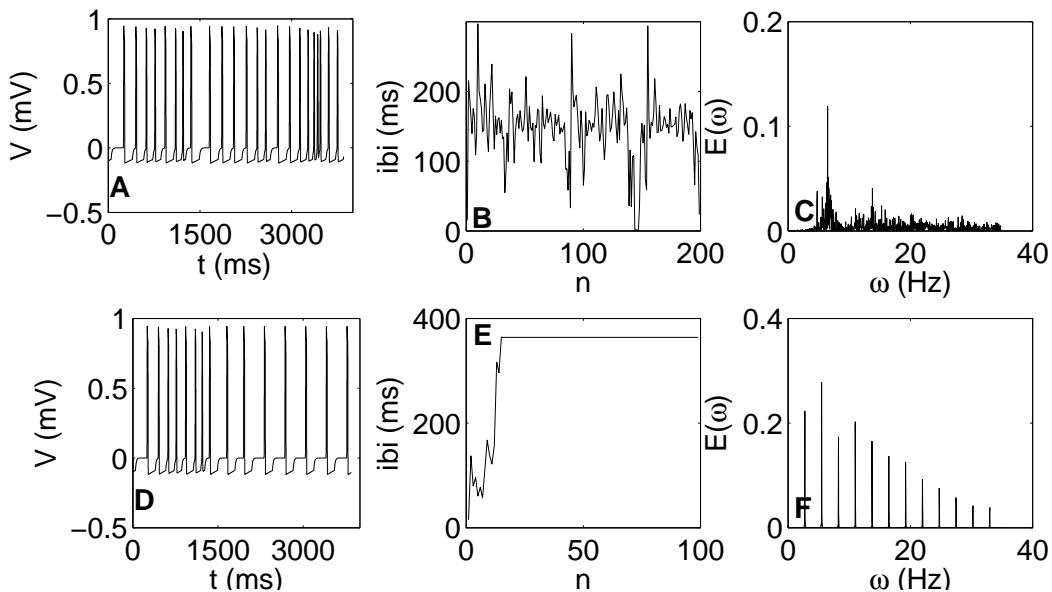}
\caption{
The local time series, interbeat interval IBI,
and power spectrum of the transmembrane potential $V(x,y,t)$ at a 
representative point $(x,y)$ in the tissue. When the obstacle is at 
(160 mm,100 mm) a spiral turbulent state ST is obtained with the time series
({\bf A}), and interbeat interval ({\bf B})showing non-periodic chaotic 
behavior and a broad-band power spectrum ({\bf C}). However, with the
bottom-left corner of the obstacle at (150 mm,100 mm), the spiral 
wave gets attached to the obstacle after 9 rotations ($\simeq$ 1800 ms);
this is reflected in the time series ({\bf D}) and the plot of the 
interbeat interval({\bf E}); after transients the latter settles on to 
a constant value of 363 ms; the power spectrum ({\bf F}) shows discrete 
peaks with a fundamental frequency $\omega_f$ = 2.74 Hz and its harmonics.
Initial transients over the first 50,000 $\delta t$ were removed
before we collected data for calculating the power spectrum.
}
\end{figure*}
To understand in detail how the position of the obstacle changes the final
state, we now present the results of our extensive simulations for the $d$ = 2
Panfilov model in a square domain with side 200 mm, i.e., 400 $\times$ 400
grid points, and with a square obstacle of side $l$=40 mm. Figure 3A shows
our simulation domain divided into small squares of side $l_p$ mm 
($l_p$=10 mm in Fig. 3A). The color of each small square indicates the final
state of the system when the position of the lower-left corner of the obstacle
coincides with that of the small square: white, black, and gray indicate, 
respectively, ST, RS, and Q. In Figs. 3B and 3C we show the rich, fractal-like
structure of the interfaces between the ST, RS, and Q regions by zooming in
successively on small subdomains encompassing sections of these interfaces 
(white boundaries in Figs. 3A and 3B) and reducing the sizes of the small
squares into which we divide the subdomain. Clearly very small changes in
the position of the obstacle can change the state of the system from ST to
Q or RS, i.e., the spatiotemporal evolution of the transmembrane potential
depends very sensitively on the position of the obstacle.

\begin{figure*}
\centering
\includegraphics[width=14cm]{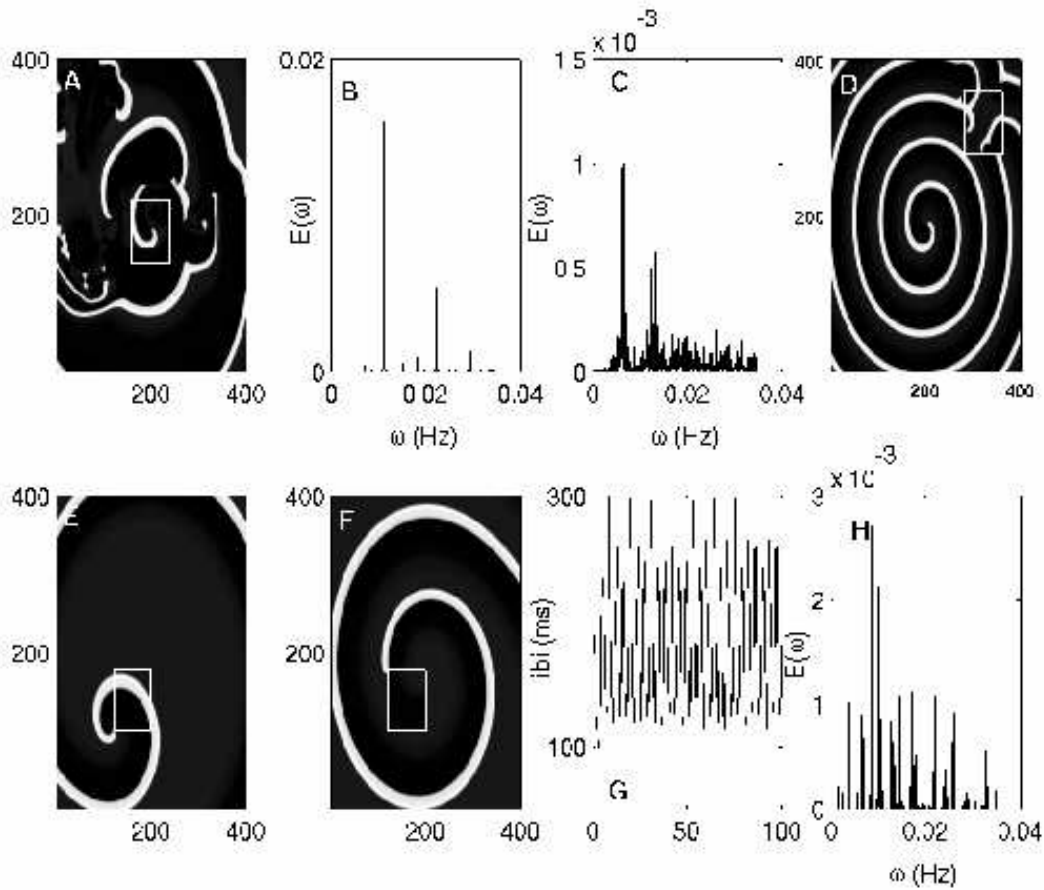}
\caption{
{\bf Inhomogeneities in $\epsilon_1$ :} Inhomogeneities in the parameter
$\epsilon_1$ result in the coexistence of different types of spatiotemporal
behavior in the same system. With $\epsilon_{1}^{out}$=0.01 and 
$\epsilon_{1}^{in}$=0.02 (see text), we obtain spatiotemporal chaos 
outside the inhomogeneity but quasiperiodic behavior inside it ({\bf A});
the latter is illustrated by the power spectrum of $V(x,y,t)$ with discrete
peaks ({\bf B}) and the former by a broad-band power spectrum ({\bf C}).
With $\epsilon_{1}^{out}$=0.03 and $\epsilon_{1}^{in}$=0.01 and the 
left-bottom corner of the inhomogeneity placed at ($x$=140 mm, $y$=140 mm),
single and broken spiral waves coexist in same medium ({\bf D}), whereas,
with the inhomogeneity at ($x$=60 mm, $y$=50 mm), a single rotating spiral 
gets anchored to the inhomogeneity ({\bf E, F}) with quasiperiodic behavior 
illustrated by the interbeat interval ({\bf G}) and the power spectrum 
({\bf H}). The power spectrum ({\bf H}) shows six frequencies 
($\omega_1 = 4.06, \omega_2 = 5.56, \omega_3 = 6.57, \omega_4 = 7.05, 
\omega_5 = 8.58$, and $\omega_6 = 9.07$ Hz) not rationally 
related to each other; all other frequencies can be expressed as
$\sum_{i=1}^6 n_i \omega_i$, where the $n_i$ are integers.
Initial transients over the first 50,000 $\delta t$ were removed
before we collected data for calculating power spectra.
}
\end{figure*}
The time series of the transmembrane potential  $V(x,y,t)$ taken from a
representative point $(x,y)$ in the simulation domain illustrates the changes
that occur when one moves from the ST to the RS regime in Fig. 3. Such time
series are shown in Fig. 4. For example, when the obstacle is placed with its
bottom-left corner at (160 mm, 100 mm), the system is in the spiral-turbulent
state ST. The time series of $V$ from the point (51 mm, 50 mm) clearly shows
non periodic, chaotic behavior. The times between successive spikes in such
time series, or interbeat intervals (IBI), are plotted versus the integers
n, which label the spikes, in Fig. 4B; this also shows the chaotic nature of
the state ST. Figure 4C shows the power spectrum $E(\omega)$ of the time
series in Fig. 4A; the broad-band nature of this power spectrum provides
additional evidence for the chaotic character of ST. By combining Figs.
4A-4C with the pseudo-greyscale plots of Figs. 1A and 1B we conclude that
ST is not merely chaotic but exhibits {\it spatiotemporal chaos}. Indeed,
it has been shown that the Panfilov model, in the spiral turbulence regime,
has several positive Lyapunov exponents whose number increases with
the size of the simulation domain; 
consequently the Kaplan-Yorke dimension also
increases with the system size (see Fig.4 of Ref.\cite{pandit});
this is a clear indication of spatiotemporal chaos.

If we change the position of the obstacle slightly and move it such that 
its left-bottom corner is at the position (150 mm, 100 mm), the spiral
eventually gets attached to the obstacle. For this case the analogs of Figs.
4A-4C are shown, respectively, in Figs. 4D-4F. From the time series of Fig.
4D we see that the transmembrane potential displays some transients
up to about 2000 ms but then it settles into periodic behavior. This is also
mirrored in the plot of IBI versus $n$ in Fig. 4E in which the transients
asymptote to a constant value for the IBI (363 ms) which is characteristic
of periodic spikes. Not surprisingly, the corresponding power spectrum in
Fig. 4F consists of discrete spikes at frequencies $\omega_m = m \omega_f$,
where $m$ is a positive integer and $\omega_f$ is the fundamental frequency 
($\omega_f$ = 2.74 Hz). A simple rotating spiral anchored at the obstacle
(Fig. 1C) will clearly result in such a periodic time series in the state RS.

We do not show the analogs of Figs. 4A-4C for the quiescent state Q since the
transmembrane potential $V$ just goes to zero after an initial period of 
transients. The durations for which the transients last, say in 
Fig. 4D, vary greatly depending on the position of the obstacle relative to the
spiral tip. We have seen transient times ranging from 300 ms to 2000 ms in our
simulations.

We obtain similar results for the three-dimensional Panfilov and the 
two-dimensional Luo-Rudy I models: Illustrative pictures from our simulations
of spiral turbulence (ST) and a single rotating spiral (RS) anchored at the 
obstacle are shown in Figs. 1 and 2, respectively. From these and similar 
figures we note that the final state, ST, RS, or Q, depends not only on the
size of the obstacle but also on its position. 
Obstacles of different shapes,
e.g., circles, irregular shapes, and two squares separated from each other,
lead to similar results (see \texttt{www.physics.iisc.ernet.in/\~{}rahul/movies.html}
for representative movies of our simulations). 

We have also explored the 
effects inhomogeneities in parameters such as $\epsilon_1$ in Eq. (1).
In the Panfilov model, $\epsilon_{1}^{-1}$ is the recovery time-constant 
for large values of $g$ and intermediate values of $V$\cite{panf-pla}. 
As $\epsilon_1$ increases the 
absolute refractory period of the action potential decreases and this in turn 
decreases the pitch of the spiral wave (cf. Fig.3 in Ref.\cite{pandit}).  

In a homogeneous simulation domain (of size say 200 $\times$ 200 mm$^2$) 
values of $\epsilon_1 >$ 0.03 produce a single periodically rotating spiral.
As $\epsilon_1$ is lowered, e.g., if $\epsilon_1 <$ 0.02, quasiperiodic
behavior is seen; this is associated with the meandering of the tip of a
simple rotating spiral. Even lower values of $\epsilon_1$, say 
$\epsilon_1= 0.01$ that we have used above, lead to spatiotemporal chaos.
We now consider an inhomogeneous simulation domain in which all parameters
in the model except $\epsilon_1$  remain constant over the whole simulation
domain. We then introduce a square inhomogeneity inside which $\epsilon_1$
assumes the value $\epsilon_{1}^{in}$  and outside which it has the value 
$\epsilon_{1}^{out}$. Different choices of $\epsilon_{1}^{in}$ and 
$\epsilon_{1}^{out}$ lead to the interesting behaviors we summarize below.

With a square patch of size 40 $\times$ 40 mm$^2$, 
$\epsilon_{1}^{in}$ = 0.02, and $\epsilon_{1}^{out}$ = 0.01, a spatiotemporally
chaotic state is obtained for most positions of this inhomogeneity. But
there are certain critical positions of this inhomogeneity for which all 
spirals are completely eliminated (e.g., when the left-bottom corner of the
inhomogeneity is at $x$=70 mm, $y$=120 mm the spiral moves out of the simulation
domain). For yet other positions of the inhomogeneity, spatiotemporal chaos
is obtained outside the inhomogeneity but inside it quasiperiodic behavior 
is seen (Figs. 5A-5C). However, with $\epsilon_{1}^{in}$ = 0.01 and
$\epsilon_{1}^{out}$ = 0.03, spiral breakup occurs inside the inhomogeneity
and coexists with unbroken periodic spiral waves outside it (Fig. 5D), as
previously noted by Xie {\it et al} \cite{xie2}. 
Even in this case, for certain positions
of the inhomogeneity, a single spiral wave gets anchored to it (Figs. 5E, 5F)
as in the case of a conduction inhomogeneity (Fig. 1C). However, the temporal
evolution of $V$ at a representative point in Fig. 5E is richer than it is in
Fig. 1C: $V(x,y,t)$, with $x$=51 and $y$=50, displays the interbeat interval of
Fig. 5G; the associated power spectrum shows six fundamental frequencies, not
rationally related to each other, and their combinations; this indicates strong
quasiperiodicity of $V(x,y,t)$.
So, even an inhomogeneity in the excitability
of the medium can cause the ST-RS or ST-Q transitions 
we have discussed above for the case of conduction inhomogeneities.
Furthermore, an inhomogeneity in excitability can also lead to 
rich temporal behaviors as shown in Figs. 5 E-H.

\section{Discussion}

We have shown that spiral turbulence in models of cardiac arrhythmias
depends sensitively on the size and position of inhomogeneities in 
the medium. In particular, we have shown that, with the inhomogeneity
at a particular position, the state of the spiral wave 
changes from ST to RS as the size of the obstacle increases. We 
have also shown that, for an obstacle 
with fixed size, this transition also depend upon the position of the obstacle.
Two important questions arise from our work: (1) What causes the sensitive
dependence of such spiral turbulence on the positions and sizes of 
conduction inhomogeneities? (2) What are the implications of our theoretical
study for cardiac arrhythmias and their control? We discuss both these 
questions below.

Spiral turbulence (ST) and a single rotating spiral (RS) in our models are
like VF and VT, respectively, in cardiac tissue.
Our study suggests, therefore, that 
such cardiac arrhythmias, like
their ST and RS analogs in the Panfilov and Luo-Rudy I models, must depend
sensitively on the positions and sizes of conduction inhomogeneities.
Furthermore, our work indicates that this is a natural consequence of the
spatiotemporal chaos associated with spiral turbulence 
\cite{pandit,sinha2} in these 
models: Even for much simpler, low-dimensional dynamical systems it is often
the case that a {\it fractal} basin boundary \cite{lai,sommerer} 
separates the basin of
attraction of a strange attractor from the basin of attraction of a fixed
point or limit cycle; thus a small change in the initial condition can
lead either to chaos, associated with the strange attractor, or to the 
simple dynamical behaviors associated with fixed points or limit cycles.

The PDEs we consider here are infinite-dimensional dynamical systems; the
complete basin boundaries for these are not easy to determine; however,
it is reasonable to assume that a complex, fractal-like boundary separate the basins of
attraction of spatiotemporally chaotic states (e.g., ST) and those with
simpler behaviors (e.g., RS or Q). Here we do not change the initial
condition; instead we change the dynamical system slightly by moving the
position, size, or shape of a conduction inhomogeneity. This too affects
the long-time evolution of the system as sensitively as does a change in
the initial conditions.

In particular, our work elucidates that, by changing the position of a
conduction inhomogeneity, we may convert spiral breakup to single rotating spiral or vice versa as depicted
graphically in Figs. 3 and 4. Even more exciting is the possibility that,
at the boundary between these two types of behavior (Fig. 3), we can find the
quiescent state Q. Thus our model study obtains all the analogs of 
possible qualitative
behaviors found in experiments, namely, (1) ST might persist even in the
presence of an obstacle, (2) it might be suppressed partially and become RS,
or (3) it might be eliminated completely.

Our work on inhomogeneities in the parameter $\epsilon_1$ 
in the Panfilov model illustrates the complex way in which the
spatiotemporal evolution of the transmembrane potential depends on the properties of
this model for cardiac tissue. 

The implications of our results for anti-tachycardia-pacing and
defibrillation algorithms, used for the suppression of cardiac arrhythmias,
are very important. Optimal pacing algorithms might well have to be tailor
made for different inhomogeneities. Indeed, clinicians often adapt their
hospital procedures for the treatment of arrhythmias, on a case-by-case basis,
to account for cardiac structural variations between patients 
\cite{christini}. We hope, therefore, that our work will stimulate 
further systematic experiments on the effects of obstacles on 
cardiac arrhythmias.
\begin{acknowledgments}
We thank V. Nanjundiah and G. Sahoo for discussions, DST, UGC, and CSIR (India) for
 support, and SERC (IISc) for computational resources. 
\end{acknowledgments}
\appendix*
\section{The Luo-Rudy Model}

In the Luo-Rudy I (LR I) model there are six components of the ionic 
current, which are formulated mathematically in terms of Hodgkin-Huxley-type 
equations\cite{hodgkin}.
The partial differential equation for the transmembrane potential $V$ is
\begin{eqnarray}
\frac{\partial V}{\partial t}+\frac{I_{LR}}{C}&=&D\nabla^2V.
\end{eqnarray}
Here $I_{LR}$ is the instantaneous, total ionic-current density.
The subscript $LR$ denotes that we use the formulation of the
total ionic current described by the Luo-Rudy Phase I (LR1)
model \cite{luorudy}, where \(I_{LR} = I_{Na} + I_{si} + I_{K}
+ I_{K1} + I_{Kp} + I_b\), with the current densities $I_{Na}$
(fast inward $Na^+$), $I_{si}$ (slow inward), $I_{K}$ 
(slow outward {\it time-dependent} $K^+$), $I_{K_1}$
({\it time-independent} $K^+$), $I_{Kp}$ (plateau $K^+$), 
$I_b$ (total background), given by:
\begin{eqnarray*}
I_{Na} & = & G_{Na}m^3hj(V - E_{Na});\\
I_{si} & = & G_{si}df(V - E_{si});\\
I_{K} & = & G_{K}xx_i(V-E_{K});\\
I_{K_1} & = & G_{K_1}K_{1\infty}(V-E_{K_1});\\
I_{Kp} & = & G_{Kp} K_p(V-E_{Kp});\\
I_b & = & 0.03921(V+59.87);
\end{eqnarray*}
and $K_{1 \infty}$ is the steady-state value of the gating variable $K_1$. 
All current densities are in units of $\mu$A/cm$^2$, voltages are in mV, and
$G_{\xi}$ and $E_{\xi}$ are, respectively, the ion-channel conductance and
reversal potential for the channel $\xi$.  The ionic currents are determined by the time-dependent ion-channel
gating variables $h$, $j$, $m$, $d$, $f$, $x$, $x_i$, $K_p$ and $K_1$ generically denoted by $\xi$, which follow ordinary differential equations of the type
\begin{eqnarray*}
\frac{d\xi}{dt} & = & \frac{\xi_{\infty}-\xi}{\tau_{\xi}},
\end{eqnarray*}
where $\xi_{\infty}=\alpha_{\xi}/(\alpha_{\xi}+\beta_{\xi})$ is
the steady-state value of $\xi$ and
$\tau_{\xi}=\frac{1}{\alpha_{\xi}+\beta_{\xi}}$ is its time constant.
The voltage-dependent rate constants, $\alpha_{\xi}$ and $\beta_{\xi}$,
are given by the following empirical equations:
\begin{eqnarray*}
\alpha_{h} & = & 0,~{\rm if}~V \geq -40~{\rm mV},\\
  & = & 0.135~\exp{[-0.147~(V + 80)]}, ~{\rm otherwise};\\
\beta_{h} & = & {\frac{1}{0.13~(1 + \exp{[-0.09 (V + 10.66)]})}},~{\rm if}~V \geq -40~{\rm mV},\\
                & = & 3.56~\exp{[0.079~V]}+3.1 \times 10^5 ~ \exp{[0.35~V]}, ~{\rm
         otherwise};\\
\end{eqnarray*}
\begin{eqnarray*}
\alpha_{j} & = & 0,~{\rm if}~V \geq -40~{\rm mV},\\
          & = &{[\frac{(\exp{[0.2444~V]}+ 2.732 \times 10^{-10}~\exp{[-0.04391~V]})}
             {-7.865 \times 10^{-6} \{1 + \exp{[0.311~(V+79.23)]} \} }]}\\
             & & \times (V + 37.78), ~{\rm otherwise};\\
\beta_{j}&=&{\frac{0.3~\exp{[-2.535 \times 10^{-7}~V]}}{1+\exp{[-0.1~(V + 32)]}}},
{\rm if}~V \geq -40~{\rm mV},\\
        &=&{\frac{0.1212~\exp{[-0.01052~V]}} {1 + \exp{[-0.1378~
(V + 40.14)]}}}, ~{\rm otherwise};\\
\end{eqnarray*}
\begin{eqnarray*}
\alpha_{m}&=&{\frac{0.32~(V+47.13)}{1-\exp{[-0.1~(V+47.13)]}}};\\
\beta_{m}&=&0.08~\exp{[-0.0909~V]};\\
\end{eqnarray*}
\begin{eqnarray*}
\alpha_{d}&=&{\frac{0.095~\exp{[-0.01~(V-5)]}}{1+\exp{[-0.072~(V-5)]}}};\\
\beta_{d}&=&{\frac{0.07~\exp{[-0.017~(V+44)]}}{1+\exp{[0.05~(V+44)]}}};\\
\end{eqnarray*}
\begin{eqnarray*}
\alpha_{f}&=&{\frac{0.012~\exp{[-0.008~(V+28)]}}{1+\exp{[0.15~(V+28)]}}};\\
\beta_{f}&=&{\frac{0.0065~\exp{[-0.02~(V+30)]}}{1+\exp{[-0.2~(V+30)]}}};\\
\end{eqnarray*}
\begin{eqnarray*}
\alpha_{x}&=&{\frac{0.0005~\exp{[0.083~(V+50)]}}{1+\exp{[0.057~(V+50)]}}};\\
\beta_{x}&=&{\frac{0.0013~\exp{[-0.06~(V+20)]}}{1+\exp{[-0.04~(V+20)]}}};\\
\end{eqnarray*}
\begin{eqnarray*}
\alpha_{K1}&=&{\frac{1.02}{1+\exp{[0.2385~(V-E_{K1}-59.215)]}}};\\
\beta_{K1}&=&{\frac{[0.49124~\exp{[0.08032~(V-E_{K1}+5.476)]}}
{1+\exp{[-0.5143~(V-E_{K1}+4.753)]}}}\\
       & &       + \exp{[0.06175~(V-E_{K1}-594.31]}].\\
\end{eqnarray*}
The gating variables $x_i$ and $K_p$ are given by
\begin{eqnarray}
\nonumber
x_i&=&{\frac{2.837~\exp{0.04 (V+77)}-1}{(V+77)\exp{0.04~(V+35)}}}, {\rm if}~V
>
-100 {\rm mV},\\
   &=&1, ~{\rm otherwise}; \\
K_p&=&{\frac{1}{1+\exp{[0.1672~(7.488-V)]}}}.
\end{eqnarray}
The values of the channel conductances $G_{Na}$, $G_{si}$, $G_K$,
$G_{K_1}$, and $G_{Kp}$ are 23, 0.07, 0.705, 0.6047 and
0.0183 mS/cm$^2$, respectively\cite{Qu99}. The reversal potentials are $E_{Na}=54.4$ mV,
$E_K=-77$ mV, $E_{K1}=E_{Kp}=-87.26$ mV, $E_b=-59.87$ mV,
and $E_{si}=7.7-13.0287\ln Ca$,
where $Ca$ is the calcium ionic concentration satisfying
\begin{equation*}
\frac{d Ca}{dt} = -10^{-4}I_{si}+0.07(10^{-4} - Ca).
\end{equation*}
The times $t$ and $\tau_{\xi}$ are in ms; the rate
constants $\alpha_{\xi}$ and $\beta_{\xi}$ are in ms$^{-1}$.

\end{document}